\documentclass[12pt,preprint]{aastex}

\shorttitle{Dust Properties of TMC-1C}
\shortauthors{Schnee, S., Kauffmann, J., Goodman, A. \&
              Bertoldi, F.}

\begin{document}

\newcommand{\nthp}{N$_2$H$^+$(1-0)}
\newcommand{\kms}{km s$^{-1}$}

\title{The Effect of Noise in Dust Emission Maps on the Derivation of
Column Density, Temperature and Emissivity Spectral Index}

\author{S. Schnee$^{1,4}$, J. Kauffmann$^{2,3,*}$, A. Goodman$^1$ \& F. Bertoldi$^3$}

\affil{$^1$Harvard-Smithsonian Center for Astrophysics, 60 Garden
       Street, Cambridge, MA 02138 \\ 
       $^2$Max-Planck Institut f\"ur Radioastronomie, Auf dem H\"ugel
       69, 53121 Bonn, Germany \\
       $^3$Argelander Institute for Astronomy, University of Bonn, Auf
       dem H\"ugel 71, 53121 Bonn, Germany \\
       $^4$Department of Astronomy, California Institute of
       Technology, MC 105-24 Pasadena, CA 91125 \\
       $^*$present address: Harvard-Smithsonian Center for
       Astrophysics, 60 Garden Street, Cambridge, MA 02138}

\email{sschnee@cfa.harvard.edu}

\begin{abstract}	
We have mapped the central 10\arcmin$\times$10\arcmin\ of the dense
core TMC-1C at 450, 850 and 1200 \micron\ using SCUBA on the James
Clerk Maxwell Telescope and MAMBO on the IRAM 30m telescope.  We show
that although one can, in principle, use images at these wavelengths
to map the emissivity spectral index, temperature and column density
independently, noise and calibration errors would have to be less than
$\sim$2\% to accurately derive these three quantities from a set of
three emission maps.  Because our data are not this free of errors, we
use our emission maps to fit the dust temperature and column density
assuming a constant value of the emissivity spectral index and explore
the effects of noise on the derived physical parameters.  We find that
the derived extinction values for TMC-1C are large for a starless core
($\sim$80 mag $A_V$), and the derived temperatures are low ($\sim$6 K)
in the densest regions of the core, using our derived value of $\beta
= 1.8$.
\end{abstract}

\keywords{stars: formation --- dust, extinction --- submillimeter}

\section{Introduction}
	Efforts to determine the mass and temperature of starless
cores from sub-millimeter and millimeter observations are hampered by
uncertainties in the emission properties of the dust grains, such as
the emissivity spectral index.  Although in principle it should be
possible to calculate the column density of dust, the emissivity
spectral index of the dust, and the dust temperature from observations
at three or more wavelengths, in practice this has never been done for
a starless core.  A similar analysis has been done for circumstellar
disks, e.g. \citep{Beckwith91, Mannings94, Mathieu95} in which
temperature gradients, disk masses and spectral indices are
calculated.  In this paper we explore the levels of uncertainty in the
derived dust temperature ($T_d$), emissivity spectral index ($\beta$)
and column density ($N_H$) resulting from datasets of either three or
four noisy emission maps at different wavelengths.  We then apply this
analysis to the starless core TMC-1C.

	TMC-1C is a starless core in the Taurus molecular cloud, at an
approximate distance of 140 pc \citep{Kenyon94}.  It was shown that
TMC-1C is a coherent core, meaning that its velocity dispersion is
roughly constant, at slightly more than the sound speed, over a radius
of 0.1 pc \citep{Barranco98, Goodman98}.  The velocity field of TMC-1C
shows evidence of solid body rotation, at 0.3 \kms\ pc$^{-1}$
\citep{Goodman93}, and the \nthp\ spectrum reveals the signature of
sub-sonic infall \citep{Schnee05a}.  The mass derived from 450 and 850
\micron\ maps alone (13 M$_\odot$) is several times the virial mass,
and the density profile is similar to that of a Bonnor-Ebert sphere
\citep{Schnee05a}.

	Here we use data taken with SCUBA (at 450 and 850 \micron) and
MAMBO (at 1200 \micron) to make maps of the dust column density and
temperature, and to estimate a constant value for the emissivity
spectral index of TMC-1C.  Although the high signal to noise at 850
and 1200 \micron\ make this set of maps one of the best yet available
for a starless core, we show that the noise is still too high to
reliably map variations in the emissivity spectral index of TMC-1C.
 
\section{Observations} \label{OBS}
\subsection{SCUBA} \label{SCUBAOBS}
We observed a 10\arcmin$\times$10\arcmin\ region around TMC-1C using
SCUBA \citep{Holland99} on the JCMT.  Our maps, especially at 450
\micron, benefitted from exceptionally stable grade 1 weather.  We
used the standard scan-mapping mode, recording the 850 and 450
\micron\ data simultaneously \citep{Pierce-Price00, Bianchi00}.  Three
chop throw lengths of 30\arcsec, 44\arcsec, and 68\arcsec\ were used
in both the right ascension and declination directions.  The JCMT has
FWHM beams of 7.5\arcsec\ at 450 \micron\ and 14\arcsec\ at 850
\micron, which subtend diameters of 0.005 and 0.01 pc, respectively,
at the distance of Taurus.  Pointing during the observations was
typically good to 3\arcsec\ or better.  The data reduction for the
SCUBA data is described by \citet{Schnee05a}.  The absolute flux
calibration is uncertain at levels of $\sim$4\% at 850 \micron\ and
$\sim$12\% at 450 \micron.  The rms noise in the 850 \micron\ map is
8.6 mJy/beam, and 13 mJy/beam in the 450 \micron\ map, measured in
regions with no significant emission.

	The Emerson2 technique used to reconstruct SCUBA scan maps
from its component chop throws introduces false structure that can be
removed \citep{Johnstone00}.  To remove this structure, we convolved
the SCUBA images (after masking out pixels with $|S| > 5 \sigma$) with
a Gaussian of FWHM twice the size of the largest chop throw, and
subtracted this from the original image, as explained in
\citet{Reid05}.  The resulting image has fluxes nearly identical to
the original in regions of high signal to noise, but has fewer bowls
of negative emission and other artifacts introduced by chopping and
image reconstruction.  We only use data from the high signal to noise
central region of the maps to estimate the dust properties.  We
account for the SCUBA error beams by convolving the 850 and 1200
\micron\ maps with the 450 \micron\ PSF, and convolving the 450 and
1200 \micron\ maps with the 850 \micron\ PSF before regridding to a
common resolution, as explained in detail in \citet{Reid05}.

	The original bowls of negative flux in our 450 \micron\ map on
either side of region `a' in Figure \ref{TMC1CFIG} have peak values
around $-150$ mJy, and average values around $-75$ mJy.  We believe
that these structures are artifacts of image reconstruction.  It is
possible that structures of similar magnitude (either positive or
negative) might be affecting data elsewhere in the map.

\subsection{MAMBO} \label{MAMBOOBS}

	We observed the 1.2 mm continuum emission in November 8, 2002,
October 23, 2003, and November 2, 2003 using the 117-channel MAMBO-2
array \citep{Kreysa99} at the IRAM 30-meter telescope on Pico Veleta
(Spain).  The FWHM beam size on the sky was 10\farcs7.  The source was
mapped on-the-fly, with the telescope subreflector chopping in azimuth
by 60\arcsec\ to 70\arcsec\ at a rate of 2 Hz; the total on-target
observing time was about 6 hours.  The line-of-sight optical depth
varied between 0.1 and 0.5.  The data were reconstructed using the EKH
algorithm in an iterative way that properly reproduces large-scale
emission (Kauffmann et al., in prep.).  The rms noise in the 1200
\micron\ map is 3 mJy/beam, measured in regions with no significant
emission.  The flux calibration uncertainty is approximately 10\%,
which is derived from the rms of calibrator observations across pool
observing sessions and the uncertainty in the intrinsic calibrator
fluxes.

\section{Solving for physical parameters} \label{EQSOLVE}

In principal, one can use three measured quantities (e.g. the three
flux density maps at 450, 850 and 1200 \micron) to solve for three
unknowns (e.g. maps of the dust temperature, emissivity spectral
index, and column density).  Our maps originally are in units of flux
density per beam, with a different beam size for each map.  To derive
meaningful physical quantities we smooth and rebin our maps to a
common resolution of the largest beam, which in our case is 14\arcsec.
As a result, the quanity that we work with is flux density per
14\arcsec\ pixel, which is what we present in our maps of TMC-1C.

	The flux density per beam in each map is given by:
\begin{equation} \label{DUSTFLUX4}
S_{\nu} = \Omega B_{\nu}(T_d) \kappa_{_{\nu}} \mu m_H N_{H_2},
\end{equation}
where
\begin{equation} \label{BLACKBODY4}
B_\nu(T_d) = \frac{2h\nu^3}{c^2} \frac{1}{\exp(h\nu /kT_d)-1}
\end{equation}
and 
\begin{equation} \label{KAPPA4}
\kappa_{_{\nu}} = \kappa_{230} \left(\frac{\nu}{230 {\rm GHz}} 
	       \right)^{\beta}.
\end{equation}
In Equation \ref{DUSTFLUX4}, $S_{\nu}$ is the flux density per
14\arcsec\ pixel; $\Omega$ is the solid angle of the beam;
$B_{\nu}(T_d)$ is the blackbody emission from the dust at temperature
$T_d$; $\kappa_{230} = 0.009$ cm$^2$ g$^{-1}$ is the emissivity of the
dust grains at 230 GHz \citep{Ossenkopf94}; $m_H$ is the mass of the
hydrogen atom; $\mu = 2.8$ is the mean molecular weight of
interstellar material in a molecular cloud per hydrogen molecule;
$N_{H_2}$ is the column density of hydrogen molecules and a
gas-to-dust ratio of 100 is assumed.  It should be noted that
$\kappa_{230}$ is uncertain by a factor of $\sim$2, and that we
assumed that $\kappa_{230} = 0.005$ and $\mu = 2.33$ (which is the
mean molecular weight per free particle for an abundance ratio of
$N(H)/N(He) = 10$ and negligible metals) in \citet{Schnee05a}.

	The ratio of two fluxes, because of the common beam size, can
be simply expressed as:
\begin{equation} \label{RATIOEQ}
\frac{S_{\nu_1}}{S_{\nu_2}} = \left(\frac{\nu_1}{\nu_2}\right)^{3+\beta}
     \left(\frac{\exp[h\nu_2/kT_d] - 1}{\exp[h\nu_1/kT_d] - 1}\right)
\end{equation}

	The dust temperature can be found independently of the dust
emissivity spectral index by taking the difference between the ratio
of fluxes, if we assume that each line of sight through the core can
be characterized by a single temperature and emissivity spectral
index:
\begin{eqnarray} \label{TEMPEQ}
\log\left(\frac{S_{450}}{S_{850}}\right) 
\log\left(\frac{1200 \micron}{850 \micron}\right) - \nonumber \\
\log\left(\frac{S_{850}}{S_{1200}}\right) 
\log\left(\frac{850 \micron}{450 \micron}\right) = \nonumber \\
\log\left(\frac{\exp\left[\lambda_T / 850 \micron \right] - 1}
               {\exp\left[\lambda_T / 450 \micron \right] - 1}\right) 
\log\left(\frac{1200 \micron}{850 \micron}\right) - \nonumber \\
\log\left(\frac{\exp\left[\lambda_T / 1200 \micron \right] - 1}
               {\exp\left[\lambda_T / 850 \micron \right]  - 1}\right) 
\log\left(\frac{850 \micron}{450 \micron}\right), 
\end{eqnarray}
where $\lambda_T = hc/kT_d$.

	Once the dust temperature is determined, the emissivity
spectral index can be calculated by:
\begin{eqnarray} \label{BETAEQ}
\beta =\log\left(\frac{S_{850}}{S_{1200}}
       \frac{\exp\left[\lambda_T / 850  \micron \right] - 1}
            {\exp\left[\lambda_T / 1200 \micron \right] - 1}\right)
       \left/
       \log\left(\frac{1200 \micron}{850 \micron}\right) \right. - 3.
\end{eqnarray}

The column density of dust can be derived from the flux at a single
wavelength (e.g. at 1200 \micron), the temperature of the dust and the
emissivity spectral index of the dust using Equation
\ref{DUSTFLUX4}.  The equivalent visual extinction can be calculated
from the column density $N_H$ using:
\begin{equation} \label{AVEQ}
A_V = N_H R_V \frac{E(B-V)}{N_H}
\end{equation}
where $N_H = 2 \times N_{H_2}$, $N_H/E(B-V) = 5.8\times10^{21}$
cm$^{-2}$ mag$^{-1}$ is the conversion between column density of
hydrogen nuclei (for our assumed gas to dust ratio) and the selective
absorption, and $R_V = A_V/E(B-V) = 3.1$ is the total to selective
extinction ratio for the low-density lines of sight similar to those
for which $N_H/E(B-V)$ has been measured \citep{Mathis90, Bohlin78}.
Although we use a constant value of $R_V$, we recognize that this
value is uncertain to within a factor of 2 between regions of high and
low column density \citep{Mathis90}, and the relation between
extinction and column density may be different for dense cores like
TMC-1C.

\section{Error Analysis} \label{ERROR}

In order to understand how the noise and reconstruction artifacts in
our emission maps will affect the accuracy of our derived dust
temperature, emissivity spectral index and column density we have run
a variety of Monte Carlo simulations.  We compare the effects of using
maps at three particular wavelengths to solve for all three parameters
to using the three measurements to fit for two parameters while
assuming a fixed value for the third.  We also show the improvements
brought about by using a fourth wavelength and fitting for all three
physical parameters.

\subsection{Illustrative Examples} \label{ERRORSOLVE}

\begin{figure}
\centerline{\includegraphics[totalheight=0.8\textheight]{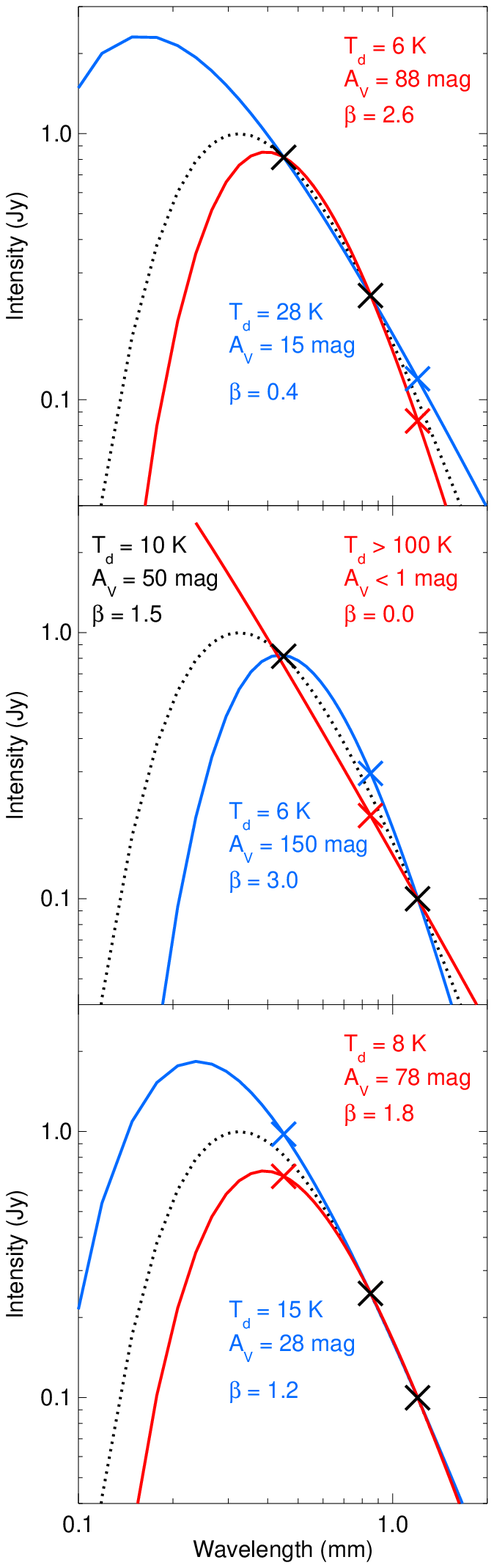}}
\caption{The true and calculated modified blackbody spectra of material 
	 with $A_V = 50$, $\beta = 1.5$ and $T_d = 10$ K, observed with a 
	 14\arcsec\ beam.  The black curve shows the true spectrum, and 
	 the black crosses show the flux at 450, 850 or 1200 \micron.  The 
	 blue and red crosses show the flux overestimated or underestimated, 
	 respectively, by 20\%.  The blue and red curves show the spectrum 
	 of dust that passes through the blue/red cross and the two black 
	 crosses.  The dust temperature, emissivity spectral index and 
	 column density that would be derived from the three data points 
	 are also shown.  
	 \label{EXSPEC}}
\end{figure}

Here we discuss the uncertainties to be expected from solving for
$T_d$, $\beta$ and $N_H$ from three emission maps that have noise or
whose calibration is uncertain.  To illustrate the method, Figure
\ref{EXSPEC} shows the modified blackbody spectrum from dust with $A_V
= 50$, $T_d = 10$ K and $\beta = 1.5$, observed with a 14\arcsec\
beam.  The black dotted line shows the true emission spectrum, with
crosses at 450, 850 or 1200 \micron, and the blue/red crosses show the
flux overestimated/underestimated at one wavelength by 20\%.  The blue
and red curves are the fitted spectra that pass through the one
blue/red cross and the two black crosses.  It is clear that a 20\%
error in one measurement creates errors in all three derived
parameters that are much larger than 20\%.  The derived values of
$T_d$, $\beta$ and $N_H$ are labeled in Figure \ref{EXSPEC}.  For
convenience, we show the derived $N_H$ in units of $A_V$.

\begin{figure}
\centerline{\includegraphics[width=4.2in]{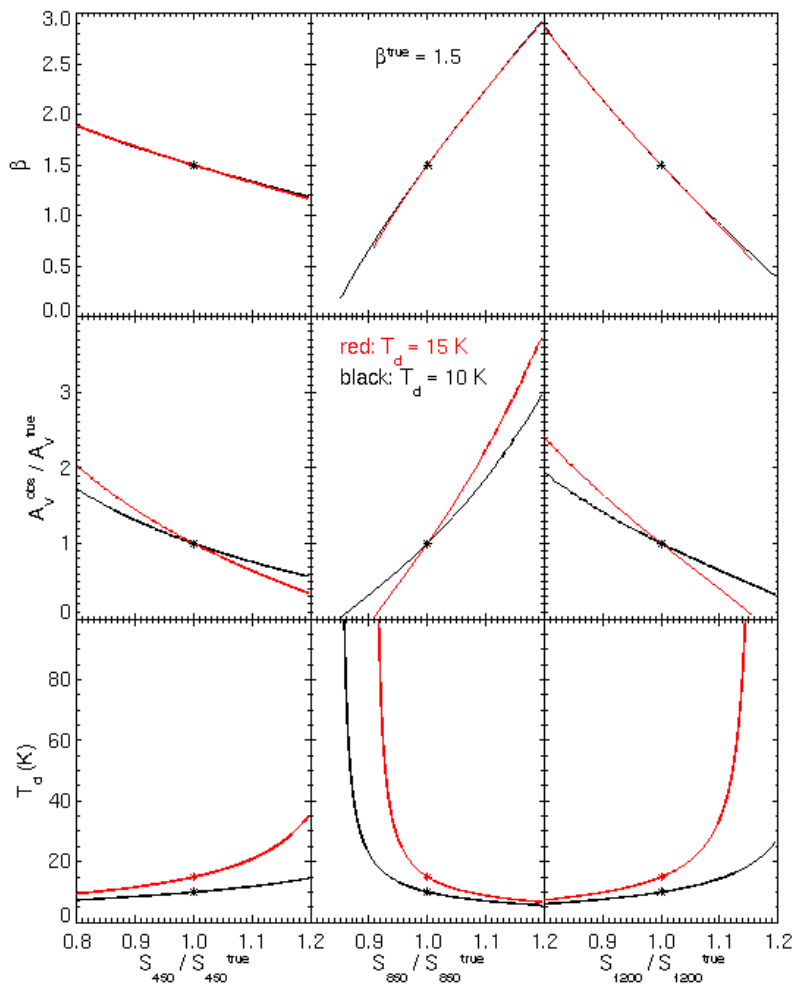}}
\caption{The derived emissivity spectral index ($\beta$), the ratio of 
	 the derived column density to the true column density and the 
	 derived temperature as functions of the ratio of the measured flux 
	 to the true flux at one wavelength (450, 850 or 1200 \micron), with 
	 the other two fluxes measured without error.  The black and red 
	 curves show the results for dust with $T_d = 10$ K and $T_d = 15$ K,
         respectively.  $\beta = 1.5$ in both cases.  The curves are cut off 
	 where the derived temperature is greater than 100 K.
	 \label{ONEFLUX}}
\end{figure}

	As can be seen from Equation \ref{TEMPEQ}, noise in any of the
three flux maps first translates into uncertainty in the derived
temperature.  This incorrect value of $T_d$, along with any noise in
the 850 and 1200 \micron\ maps then results in uncertainties in the
derived values of $\beta$ and $N_H$.  From Equation \ref{BETAEQ}, one
can show that overestimating the temperature will result in an
underestimate of $\beta$, and underestimating the temperature will
result in an overestimate of $\beta$.  The errors in the derived
parameters from incorrectly measuring the flux at one wavelength,
while correctly measuring the flux at the other two wavelengths is
shown in Figure \ref{ONEFLUX}.  The anti-correlation between $T_d$ and
$\beta$ is clearly seen.  Also apparent is that for dust with
``core-like'' values of $\beta$, $T_d$ and $N_H$, errors in 450
\micron\ flux result in smaller errors in the derived physical
parameters than errors at 850 and 1200 \micron, which is convenient
because the 450 \micron\ maps often suffer from higher levels of noise
than maps at 850 and 1200 \micron.

\subsection{Deriving the Physical Parameters} \label{ERRORFIT}

\begin{figure}
\centerline{\includegraphics[width=4.0in]{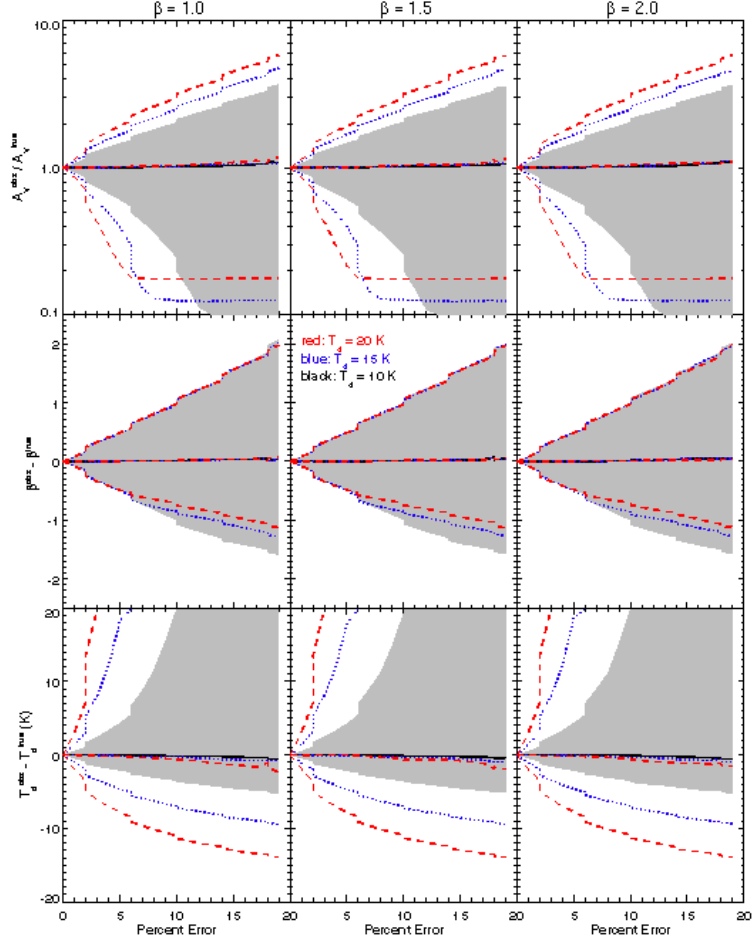}}
\caption{Each panel shows the 1$\sigma$ spread in the derived column 
	 density, temperature or emissivity spectral index as a function 
	 of the percent noise in all three flux maps at 450, 850 and 
	 1200 \micron.  A 5\% error in the measured flux corresponds to a
	 signal to noise of 20.  The black line shows the median value 
	 for the derived parameter, and the greyed area shows the 1 $\sigma$ 
	 spread in that parameter for dust at 10 K.  The blue (dotted) 
	 and red (dashed) lines show the median value and 1 $\sigma$ 
	 spread for dust at 15 and 20 K, respectively, and each panel is 
	 repeated for emissivity spectral index ($\beta$ = 1.0, 1.5 or 
	 2.0).  $N_H$, $T_d$ and $\beta$ are derived as in Section
	 \ref{EQSOLVE}.  The column density is plotted as the ratio of 
	 the derived $A_V$ to the input $A_V$, and the temperature and 
	 emissivity spectral index are plotted as the difference between 
	 the derived parameter and the true value.  
	 \label{ALLTHREE}}
\end{figure}

	In order to determine the effect of similar levels of noise in
all three wavelengths on the three derived physical parameters, we
determine the flux at 450, 850 and 1200 \micron\ from dust at $T_d =$
10, 15 and 20 K and $\beta =$ 1.0, 1.5 and 2.0.  The fluxes are then
modified by a multiplicative factor $f = 1.0 + \delta$ where
$\delta$ is randomly chosen from a normal distribution of mean zero
and standard deviation $\sigma$, and each flux is modified by a
different $f$.  This is repeated 10,000 times for each value of
$\sigma$ between 0 and 0.2.  We show the effect of noise in all three
wavelengths on the derived column density, temperature and emissivity
spectral index in Figure \ref{ALLTHREE}.  At a signal to noise of 20
(5\% error), the expected uncertainties in the derived $N_H$, $T_d$
and $\beta$ are approximately 50\%, 80\% and 40\%, respectively, for
15 K dust with $\beta = 1.5$.  The median values of the derived
parameters stay close to the input values at every value of $T_d$ and
$\beta$ tested.

	The analysis presented here deals with the special case that
the signal to noise is wavelength independent.  In general, this will
not be the case, and one might expect that for a given amount of
observing time the signal to noise will be worse at 450 \micron\ than
at 850 or 1200 \micron\ due to atmospheric effects.  In addition, the
relative signal to noise between maps will change from position to
position, due to the gradients in the dust temperature, column density
and emissivity spectral index.  For instance, at the position of the
column density peak in TMC-1C, the signal to noise ratios at 450, 850
and 1200 \micron\ are 5.3, 27 and 33, respectively, when including
both the random noise and the artifacts in the 450 \micron\ image.

\begin{figure}
\centerline{\includegraphics[width=4.5in]{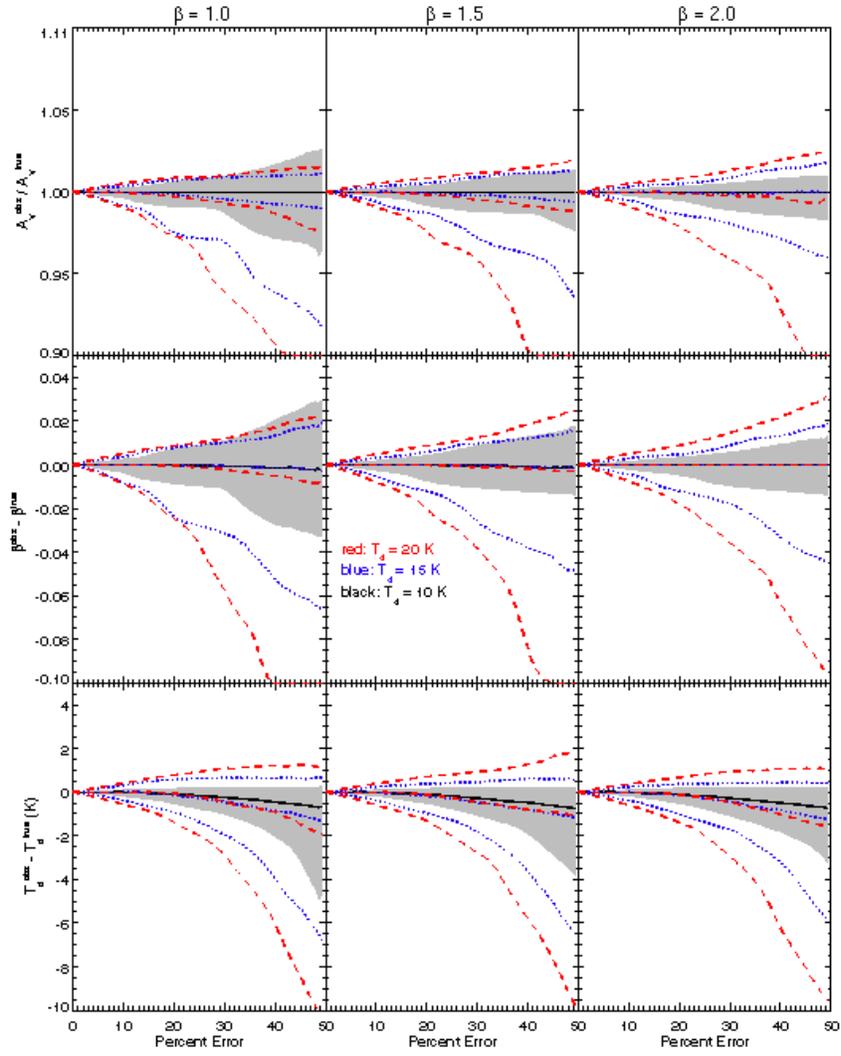}}
\caption{Same as Figure \ref{ALLTHREE}, except $N_H$, $T_d$ and $\beta$
         are fit from {\it four} observations at 350, 450, 850 and 1200 
	 \micron.
	 \label{ALLFOUR}}
\end{figure}

Although even just 5\% errors in the measured fluxes at three
wavelengths make accurate determinations of three physical parameters
impossible, adding a fourth wavelength (for instance, at 350 \micron\
or 2.7 mm) drastically reduces the effects of noise, as can be seen by
comparing Figure \ref{ALLTHREE} with Figure \ref{ALLFOUR}.  To
properly constrain the dust parameters, at least one of the
observations should not be on the Rayleigh-Jeans portion of the
emission spectrum, or else $T_d$ and $N_H$ will be degenerate in the
fit.  For dust at $T_d =$ 10, 15 and 20 K and $\beta$ = 1.0, 1.5 and
2.0 we calculate the flux at 350, 450, 850 and 1200 \micron.  As
before, the fluxes are then modified by a multiplicative factor $f =
1.0 + \delta$, where $\delta$ is again a variable randomly chosen from
a normal distribution with mean zero and standard deviation $\sigma$.
This is repeated 10,000 times for each value of $\sigma$ between 0 and
0.5.  We allow $\sigma$ to be larger than in the previous Monte Carlo
simulation because the effects of noise are smaller in this case.  For
each set of four fluxes, the column density, temperature and
emissivity spectral index are fit and the results are shown in Figure
\ref{ALLFOUR}.  At a signal to noise of 20 (5\% error), the expected
errors in the column density, temperature and emissivity spectral
index are all on the order of $\sim$1\%.  As the signal to noise gets
lower, the median derived temperature and column density decrease, and
this effect is larger for warmer cores.

\subsection{Fixing One Parameter} \label{FIXING}

\begin{figure}
\centerline{\includegraphics[width=4.5in]{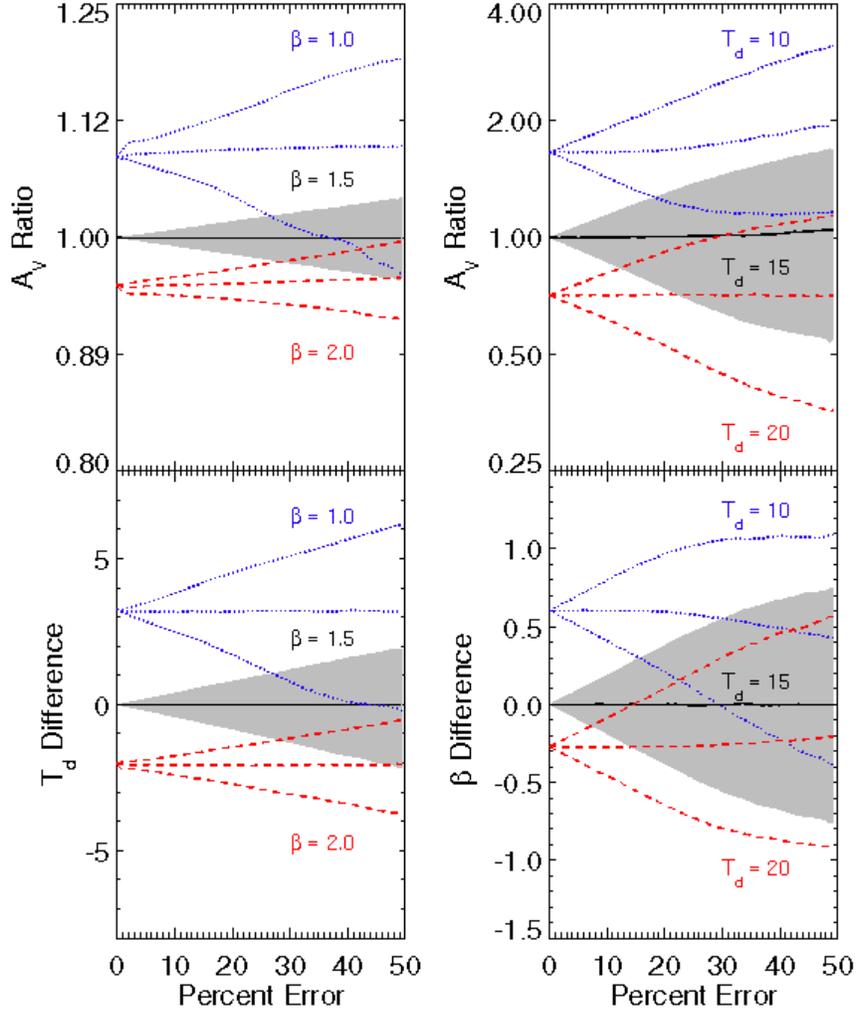}}
\caption{Each panel shows the 1 $\sigma$ spread in the derived $T_d$,
	 $\beta$ and $N_H$ for dust at $T_d = 15$ K and $\beta = 1.5$
	 from a least squares fit between the predicted and given fluxes at 
	 450, 850 and 1200 \micron.  The grey area uses the correct 
	 assumption that $\beta = 1.5$ or $T_d = 15$, while the blue 
	 (dotted) and red (dashed) lines incorrectly assume that 
	 $\beta = 1.0$ and $\beta = 2.0$, or $T_d = 10$ and $T_d = 20$, 
	 respectively.
	 \label{FITFIX}}
\end{figure}

In the case of a starless core observed at three wavelengths, one can
hold one parameter fixed, such as assuming a constant $T_d = 15$ K or
$\beta = 1.5$, and use the three flux measurements to fit the
remaining two parameters.  Figure \ref{FITFIX} shows the result of
using noisy flux measurements to fit the column density and either the
dust temperature or the emissivity spectral index of a core with $T_d
=$ 15 K and $\beta =$ 1.5.  We show the results for correctly assuming
that $\beta = 1.5$ and erroneously assuming that $\beta =$ 2.0 or 1.0,
as well as for correctly assuming that $T_d =$ 15 K and erroneously
assuming that $T_d = $ 10 or 20 K.

We see that by choosing the correct value of $\beta$ and fitting the
$T_d$ and $N_H$, noise in the emission maps on the level of 10\%
result in $\sim$1\% uncertainties in the derived temperature and
column density.  Choosing a value of $\beta = 1.0$, when the proper
value is $\beta = 1.5$ results in temperatures that are on average 3 K
too high, with a spread of $\sim$3 K, and column densities that are
high by 8\%.  Overestimating the emissivity spectral index, by
assuming that $\beta = 2.0$, results in temperatures that are on
average 2 K too low and column densities that are 5\% low.

Figure \ref{FITFIX} shows that by choosing the $T_d$ and fitting for
the $\beta$ and $N_H$, the true values of $\beta$ and column density
are recovered, on average, with an absolute spread of 0.20 in $\beta$
and a relative spread of 13\% in column when the signal to noise ratio
is 10.  Underestimating the temperature by assuming $T_d = 10$ K when
the proper value is $T_d = 15$ K results in an overestimate of the
emissivity spectral index by 0.6, on average, and an overestimate of
the column density by 70\%.  Overestimating the temperature by
assuming that $T_d = 20$ K results in an underestimate of $\beta$ by
0.3, on average, and an underestimate of $N_H$ by a factor of 30\%.

\subsection{A Model Core} \label{TMC1CEX}

We apply the error analysis presented in Section
\ref{ERROR} to our data on the starless core TMC-1C in order to derive
new science beyond what was possible in our earlier paper on this core
\citep{Schnee05a}.  The random noise in the emission maps, as measured
in regions with faint emission, is found to be 13 mJy/beam, 9 mJy/beam
and 3 mJy/beam, which corresponds to S/N values of 21, 14 and 15 at
450, 850 and 1200 \micron\ at the (0,0) position of the emission maps.
Calibration uncertainties are $\sim$12\%, 4\% and 10\% at 450, 850 and
1200 \micron, respectively, and the image reconstruction artifacts in
the 450 \micron\ map peak at 150 mJy/beam.

\begin{figure}
\centerline{\includegraphics[totalheight=0.7\textheight]{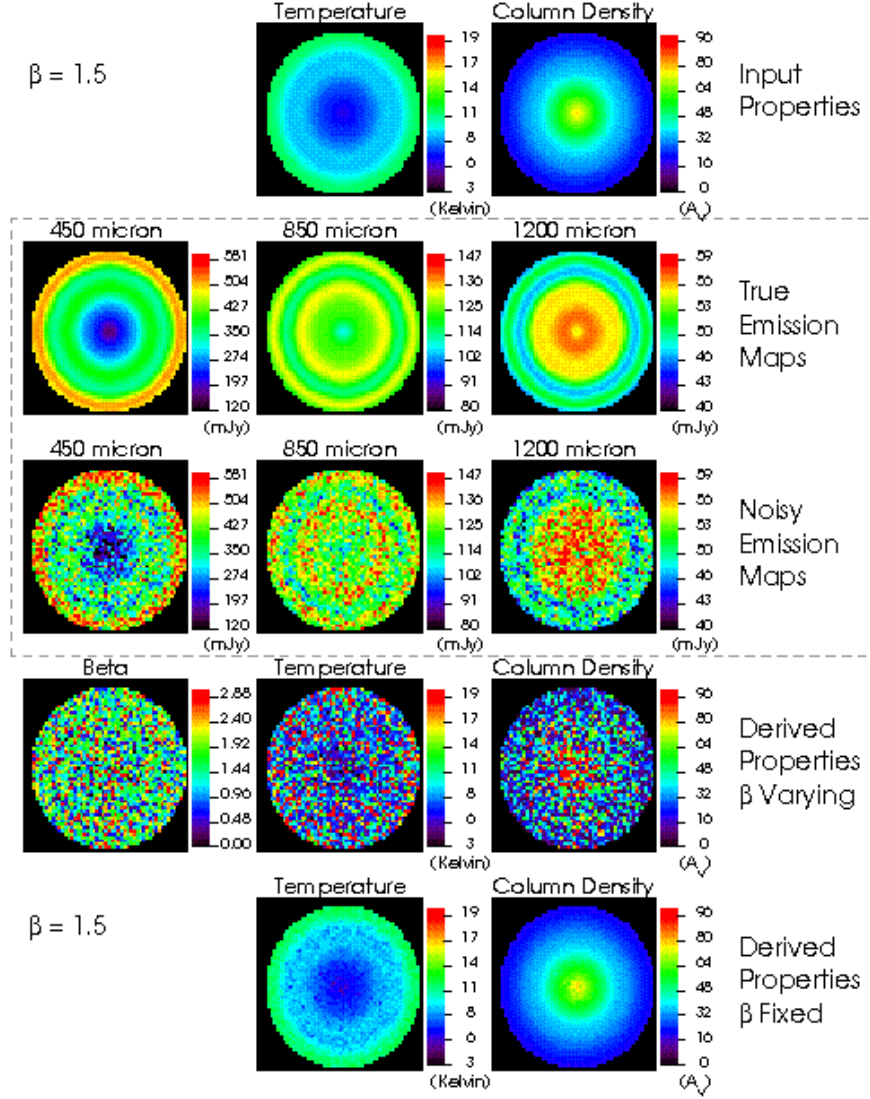}}
\caption{The top row of panels shows the temperature and column density
	 maps of a model starless core similar to the central 2\arcmin\
	 of TMC-1C.  The second row shows the predicted 
	 emission maps at 450, 850, and 1200 \micron, using a constant 
	 $\beta = 1.5$.  We then add Gaussian random noise to the 
	 emission maps equal to the measured noise and artifacts in our 
	 450, 850 and 1200 \micron\ maps of TMC-1C (80, 9 and 3 mJy, 
	 respectively) and show the resultant fluxes in the third row.  
	 Using the altered fluxes, we attempt to solve for $\beta$, $T_d$ 
	 and $N_H$ at each point in the map, and show the results in the 
	 fourth row from the top.  When we use the value $\beta = 1.5$ 
	 everywhere along with the altered 450, 850 and 1200 \micron\ 
	 fluxes to fit $T_d$ and $N_H$, we derive the maps shown in 
	 the bottom row.
	 \label{THREECORE}}
\end{figure}

To determine how the random noise and reconstruction artifacts (which
are spatially correlated and therefore not truly random) in the
observed emission maps affect the derived parameters, we construct
synthetic emission maps of a starless core like TMC-1C.  The model
core has a temperature and column density profile equal to the one
derived from the two dimensional temperature and column density
profile of TMC-1C, derived from fitting the 450, 850 and 1200 \micron\
data and assuming a constant $\beta = 1.5$.  The model starless core
is made of cylindrical shells seen face on with a central temperature
of $\sim$6 K, rising to $\sim$12 K at the edge.  The central column
density corresponds to $\sim$80 magnitudes of visual extinction,
falling to an $A_V$ of $\sim$20 at the edge.  Using the equations in
Section \ref{EQSOLVE}, we derive the resultant emission maps at 450,
850 and 1200 \micron, add in Gaussian noise of the same magnitude as
the random noise and reconstruction artifacts described above, and
from these emission maps derive maps of the column density,
temperature and emissivity spectral index.  The resultant maps are
shown in Figure \ref{THREECORE}.

	The dust emission in our model core (shown in Figure
\ref{THREECORE}) at 450 and 850 \micron\ is not a good tracer the dust
column density, though the 1200 \micron\ emission map does resemble
the column density distribution.  Spatial gradients in the dust
properties, such as the temperature and emissivity spectral index,
need to be taken into account when attempting to find the ``peak'' of
the dust distribution, even when using the relatively longer
wavelength 1200 \micron\ emission map as a proxy for column density.

	When we solve for all three physical parameters, we find that
the derived $\beta$ of our model cloud has a median value of 1.5
(which is the input $\beta$ everywhere in the model) with a standard
deviation of 0.7.  Given the close correspondence between the input
value of $\beta$ and the median value derived for it, we see that a
constant value for the emissivity spectral index can be estimated in
this manner.  Using this constant value for $\beta$ everywhere, we can
then use our three flux maps to fit the temperature and column
density.  The resultant maps of our model constructed in this way are
shown in Figure \ref{THREECORE}.  Although using the median value in
the $\beta$ map to derive a constant value for the emissivity spectral
index reduces the impact of statistical uncertainty due to random
noise, systematic shifts, such as those created from calibration
uncertainties, are not removed.
	
\section{Dust Emission in TMC-1C} \label{DISCUSSION}
\subsection{Morphology} \label{MORPHOLOGY} 

\begin{figure}
\centerline{\includegraphics[width=5.5in]{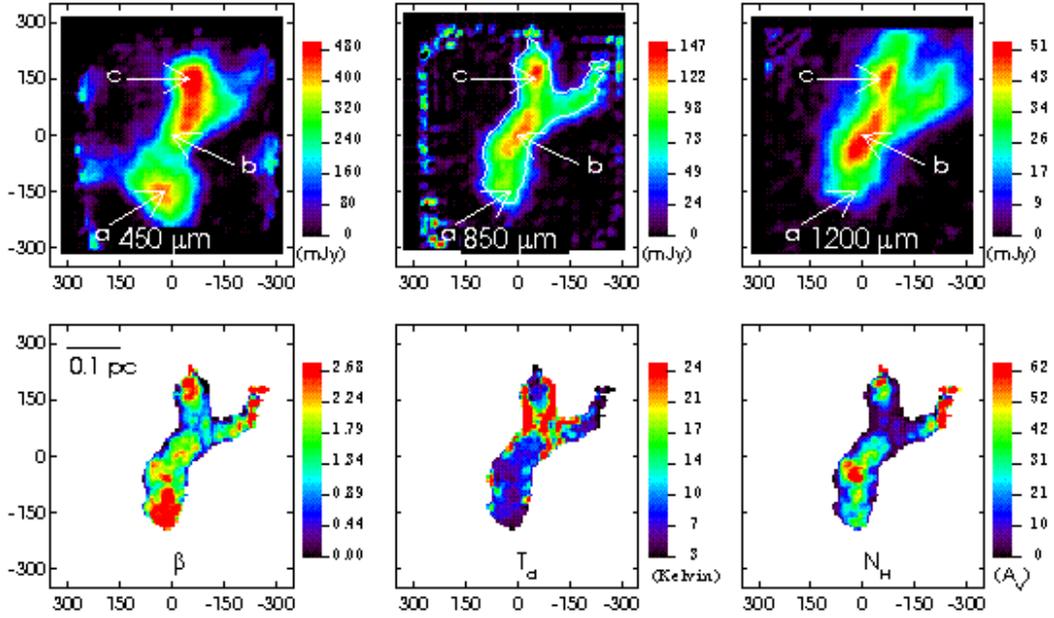}}
\caption{The 450, 850 and 1200 \micron\ emission maps of TMC-1C, along 
         with the derived $\beta$, $T_d$ and $N_H$ (expressed in units 
	 of $A_V$ for convenience).  The fluxes are in units of mJy per 
	 14\arcsec\ pixel.  The derived parameters are shown within the 
	 white 850 \micron\ contour.  The (0,0) position is at 
	 RA=4:41:35.8 DEC=+26:00:42.5 (J2000).  Position 'a' is strong 
	 at 450 \micron\ but not at 850 or 1200 \micron.  Position 'b' 
	 is strong at 850 and 1200 \micron\ but not at 450 \micron\, and 
	 position 'c' is strong in all three maps.
	 \label{TMC1CFIG}}
\end{figure}

	The observed 450 \micron\ emission map of TMC-1C is
qualitatively different from the 1200 \micron\ map.  The 450 \micron\
map shows a condensation at $(50,-150)$ (position ``a'' in Figure
\ref{TMC1CFIG}) which is much fainter at 850 \micron\ and nearly
absent at 1200 \micron.  The condensation at $(-50,150)$ (position
``c'' in Figure \ref{TMC1CFIG}) is prominent at all wavelengths, while
the column density peak at $(0,0)$ (position ``b'' in Figure
\ref{TMC1CFIG}) is prominent at 850 and 1200 \micron, but not apparent
at 450 \micron.

	An emission peak in the longer wavelength maps, but not
prominent at 450 \micron, can be explained by a cold temperature, as
seen at the $(0,0)$ position.  The 450 \micron\ emission peak not seen
at 1200 \micron\ can be explained by the dust in that region having a
steep spectral index ($\beta \ge 2$).

\subsection{Derived Parameters} \label{DERIVED4}
	
	Based on our analysis in Section \ref{ERRORSOLVE}, calibration
uncertainties, reconstruction artifacts and noise in our TMC-1C
emission maps prevent us from making accurate maps of dust
temperature, emissivity spectral index and column density {\it
simultaneously}, even though this may well be the highest S/N set of
such maps of a starless core to date.  Figure \ref{TMC1CFIG} shows the
results of an attempt to do so, and as expected the range of
temperatures that we see is quite broad ($T_d < 5$ K and $T_d > 25$ K)
for a starless core, with a similarly large spread in emissivity
spectral index ($\beta < 0$ and $\beta > 2.5$) and column density.
Furthermore, the errors and uncertainties in our emission maps create
a spurious anti-correlation between the dust temperature and
emissivity spectral index, which is also seen in our attempt to derive
$T_d$, $\beta$ and $N_H$ from our simple model of a cylindrically
symmetric starless core, described in Section \ref{TMC1CEX}.  Because
the noise and errors in our observed emission maps of TMC-1C will
drive an anti-correlation between the derived dust temperature and
emissivity spectral index even when such a trend does not exist, our
results are consistent with a constant value of the emissivity
spectral index.  However, we cannot rule out a real anti-correlation
such as that observed at 200, 260, 360 and 580 \micron\ in the Orion
molecular cloud \citep{Dupac01} and the M17 star-forming complex
\citep{Dupac02}.

\begin{figure}
\centerline{\includegraphics[width=4.0in]{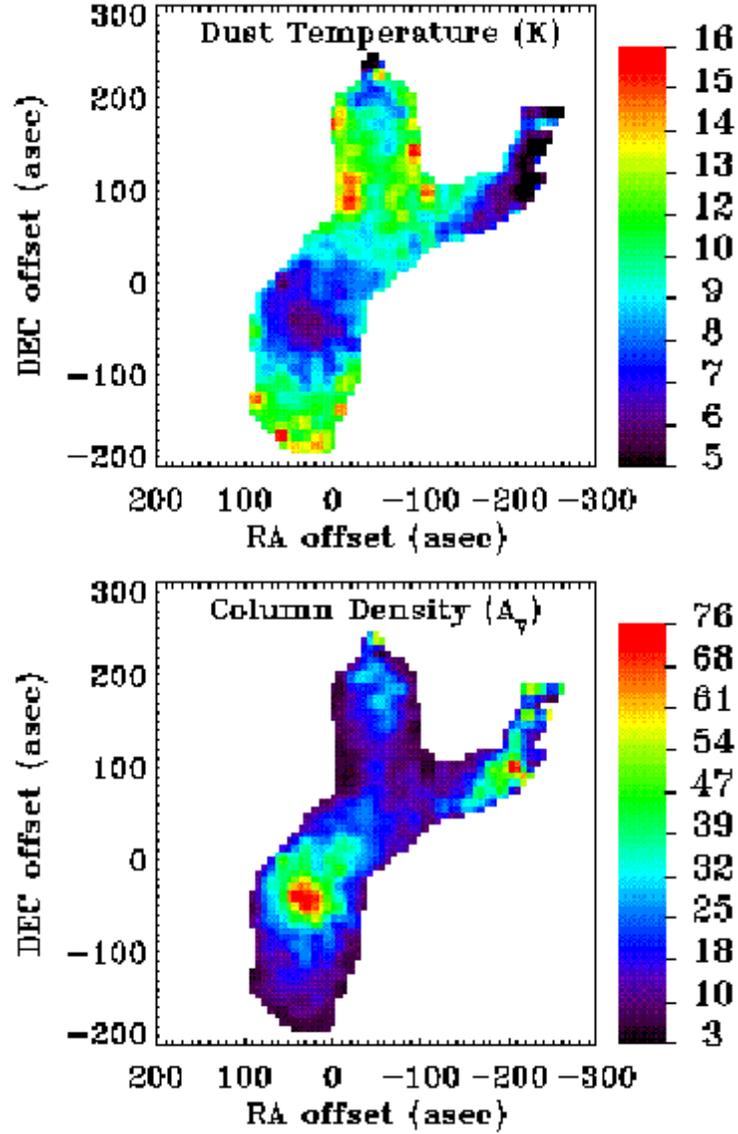}}
\caption{The TMC-1C dust color temperature (top) and column density 
	 (bottom), derived from a fit to the 450, 850 and 1200 \micron\ 
	 emission maps, assuming that the emissivity spectral index is 
	 constant at $\beta = 1.8$.
	 \label{TMC1CTDAV}}
\end{figure}

Following the method described in Section \ref{TMC1CEX}, we take the
median value of the emissivity spectral index map (Figure
\ref{TMC1CFIG}) and derive a value and uncertainty of $\beta = 1.8 \pm
0.5$ for the portion of TMC-1C located within the white contour of the
850 \micron\ image in Figure \ref{TMC1CFIG}.  Unless otherwise state,
in our subsequent analysis of TMC-1C we use a constant value of $\beta
= 1.8$ everywhere in the core.  We derive $\beta$ and the 1$\sigma$
uncertainty by running 1000 realizations of a Monte Carlo simulation
of the observed flux maps in TMC-1C modified by the calibration
uncertainties.  Using $\beta = 1.8$, we construct dust temperature and
column density maps (Figure \ref{TMC1CTDAV}) from a fit to the 450,
850 and 1200 \micron\ images.  The column density in Figure
\ref{TMC1CTDAV} peaks around the maximum of the 850 and 1200
\micron\ emission.  The implied visual extinction is quite high,
rising above 80 magnitudes in the densest regions.  As expected, the
regions with the highest column density are also the regions with the
lowest dust temperature \citep{Zucconi01}.  By using a constant value
of the emissivity spectral index ($\beta = 1.8$), the dust temperature
that we derive is nowhere significantly higher than 16 K nor lower
than 6 K.

	Our derived value of $\beta = 1.8 \pm 0.5$ is somewhat higher
than that measured for amorphous carbon grains ($\beta = 1.2$) by
\citet{Mennella98} and is within the range ($\sim 1.2 - 2.5$) measured
for silicate grains by \citet{Agladze96}.  Dust in interstellar disks
are generally observed to have values of $\beta \le 1$, but with
considerable spread \citep{Beckwith91,Mannings94}.  Observations by
\citet{Stepnik03} have shown that $\beta = 1.9 \pm 0.2$ for a dense
filament in Taurus, which agrees very well with our estimate of
TMC-1C.  Graphite and silicate dust grains in the ISM are often
assumed to have $\beta = 2$ \citep[e.g.][]{Draine84}.

\subsection{Comparison With Previous Results} \label{COMPARISON4}

	Two-dimensional temperature and column density maps of TMC-1C,
along with the deprojected three-dimensional temperature and density
profiles, have previously been reported in \citet{Schnee05a} using
only SCUBA 450 and 850 \micron\ data.  With the addition of the MAMBO
1200 \micron\ map, we are better able to constrain the temperature and
density and estimate the emissivity spectral index.  The emissivity
spectral index that we use here ($\beta = 1.8$) is higher than in
\citet{Schnee05a} ($\beta = 1.5$), because in this paper we are able
to derive $\beta$, and in our ealier work we had to assume a value.
The value that we use in this paper for $\kappa_{230}$ is taken from
\citet{Ossenkopf94}, which is larger than the number we used in
\citet{Schnee05a}.  As a result using larger $\beta$ and
$\kappa_{230}$, the temperatures that we derive are somewhat lower and
the densities are also lower.  We choose here to use the dust opacity
appropriate for dust grains with thin ice mantles, evolved in a dense
(10$^6$ cm$^{-3}$) region for 10$^5$ years derived in
\citet{Ossenkopf94} because this is the consensus value settled upon
by the {\it Spitzer} Legacy Project, ``From Molecular Cloud Cores to
Planet Forming Disks'' \citep{Evans03}, which will make comparisons
with other c2d cores easier in the future.  The mass that we derive
for TMC-1C, within a radius of 0.06 pc from the column density peak is
6 M$_\odot$, as compared with 13 M$_\odot$ in \citet{Schnee05a}.
However, even this new dust-derived mass is higher than the virial
mass derived from TMC-1C N$_2$H$^+$ observations.

\begin{figure}
\centerline{\includegraphics[width=4.0in]{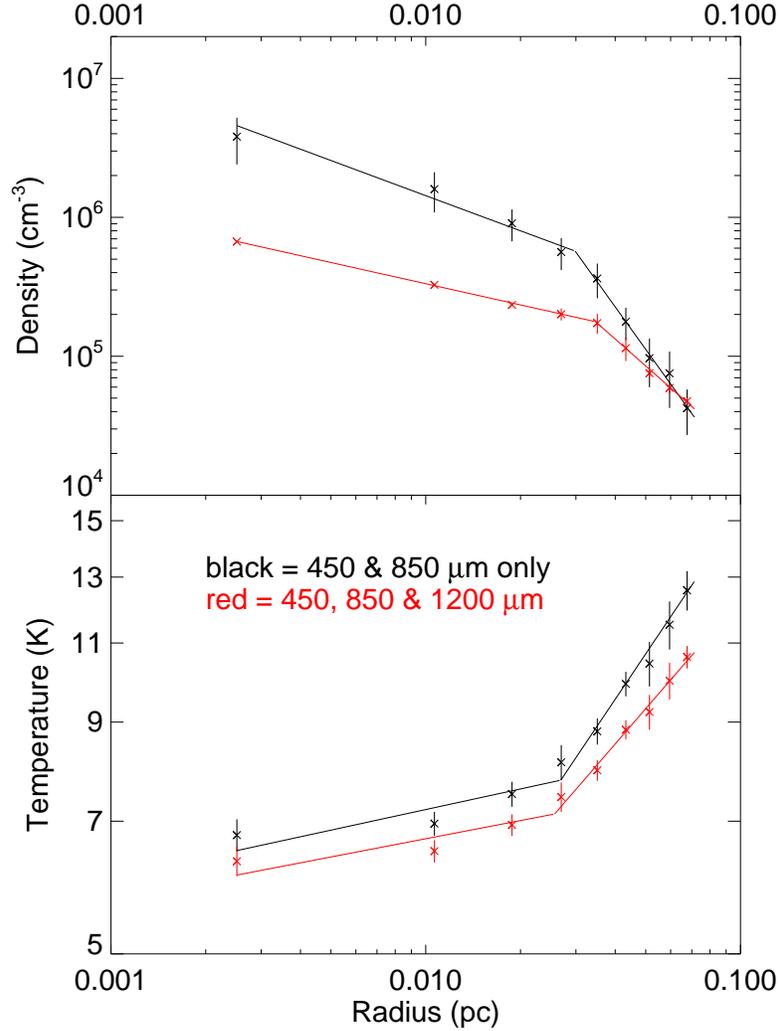}}
\caption{The density (top) and temperature (bottom) profiles of TMC-1C.  
	 The points in red are determined by a fit to the 450, 850 and 
         1200 \micron\ data, assuming $\beta = 1.8$ and $\kappa_{230} = 
	 0.009$ and the black points are determined using only the 450 
	 and 850 \micron\ data assuming $\beta = 1.5$, as in 
	 \citet{Schnee05a} and $\kappa_{230} = 0.005$, as assumed here.  
	 In both cases, we assume that the central 2 arcminute radius of 
	 TMC-1C can be modeled as nested spherical shells of uniform 
	 density and temperature.
	 \label{PROFILES}}
\end{figure}

	Following the method used in \citet{Schnee05a}, we create
deprojected three dimensional temperature and density profiles.  We
assume that the inner 0.07 pc of TMC-1C can be approximated as a set
of nested spherical shells, each with a constant temperature and
density.  The derived temperature and density profiles are shown in
Figure \ref{PROFILES}, along with the profiles calculated using only
the 450 and 850 \micron\ data and assuming that $\beta = 1.5$, as in
\citet{Schnee05a}.  The dust temperature profile that we derive is not
much changed from that derived in \citet{Schnee05a}, and is thus still
consistent with the dust temperature profiles predicted for externally
heated starless cores with Bonnor-Ebert density distributions
calculated by \citet{Evans01, Goncalves04, Stamatellos04}.  A
Bonnor-Ebert profile is a good fit to the TMC-1C density profile at
radii greater than 0.004 pc, but as in \citet{Schnee05a}, the density
we derive for the innermost point is significantly higher than
predicted by a Bonnor-Ebert model.  The density profile, shown in
Figure \ref{PROFILES}, is consistent with a broken powerlaw, with
$n(r) \propto r^{-0.5}$ inside 0.035 pc and $n(r) \propto r^{-2.0}$
outside 0.035 pc.  This is considerably flatter than the density
profile derived using only the 450 and 850 \micron\ data and $\beta =
1.5$, which have powerlaw exponents of $-0.8$ and $-3.1$ inside and
outside the break radius, respectively (also shown in Figure
\ref{PROFILES}).  In \citet{Schnee05a} we report inner and outer
powerlaw exponents of $-0.8$ and $-1.8$, also using just the 450 and
850 \micron\ data and $\beta = 1.5$, but using a slightly different
position as the center of the nested spherical shells, so the
comparison of powerlaw exponents presented in \citet{Schnee05a} and
those present here would be unfair.  Also note that the central
density of TMC-1C that we derive from the 450, 850 and 1200 \micron\
maps is a factor of $\sim$5 lower than when derived using just the 450
and 850 \micron\ maps (and assuming $\beta = 1.5$), which shows the
significant adjustments that can result from utilizing additional
emission maps.  An improved value of the central density can make a
significant impact on the predictions of the dynamical state of a
core, and on chemical models.

\section{Summary} 

We have used SCUBA data at 450 and 850 \micron\ and MAMBO data at 1200
\micron\ to create maps of the dust temperature and column density in
TMC-1C, improving the results presented in \citet{Schnee05a}.  In
addition, we are able to estimate the emissivity spectral index,
finding a value of $\beta = 1.8 \pm 0.5$, based on calibration
uncertainties.

Our analysis shows that noise and calibration errors in maps at 450,
850 and 1200 \micron\ would have to be less than $\sim$2\% to
accurately measure the dust temperature, emissivity spectral index and
column density from three emission maps.  Although such low levels of
noise and calibration uncertainties are not achievable with the
current generation of bolometers, we show that the dust temperature,
emissivity spectral index and column density can be accurately mapped
if they are fitted at four wavelengths, for instance by including the
350 \micron\ SHARCII waveband.  Obtaining accurate maps of $T_d$,
$\beta$ and $N_H$ are necessary for accurate determinations of density
and temperature profiles as well as the core mass, which determine the
time evolution and chemistry of the core.

The next generation detectors on the JCMT (SCUBA-2) and on APEX
(LABOCA), will not need to sky-chop, allowing more large-scale
structure to be visible, improved calibration and fewer image
artifacts, making more accurate determinations of dust properties in
cores possible \citep{Ellis05, Gusten06}.  

\acknowledgments 

We would like to thank Phil Myers, Ramesh Narayan, David Wilner and
Doug Johnstone for their suggestions, assistance, and insights.  The
suggestions of our anonymous referee have made substantial
improvements to this paper.  The James Clerk Maxwell Telescope is
operated by The Joint Astronomy Centre on behalf of the Particle
Physics and Astronomy Research Council of the United Kingdom, the
Netherlands Organisation for Scientific Research, and the National
Research Council of Canada.  IRAM is supported by INSU/CNRS (France),
MPG (Germany), and IGN (Spain).  This material is based upon work
supported under a National Science Foundation Graduate Research
Fellowship.

\clearpage

\clearpage

\clearpage

\clearpage

\clearpage

\clearpage

\clearpage

\clearpage

\clearpage

\end{document}